\documentclass[review,number,sort&compress]{elsarticle}
\biboptions{}

\usepackage{hyperref}
\usepackage{lineno}
\modulolinenumbers[5]

\journal{Nuclear Instruments and Methods in Physics Research}

\begin{document}

\begin{frontmatter}

\title{Cryogenic Dark Matter Search Detector Fabrication Process and Recent Improvements}

\author[mymainaddress]{A. Jastram\corref{mycorrespondingauthor}}
\cortext[mycorrespondingauthor]{Corresponding author}
\ead{akjastram@tamu.edu}

\author[mymainaddress]{H. R. Harris}
\author[mymainaddress]{R. Mahapatra}
\author[mymainaddress]{J. Phillips}
\author[mymainaddress]{M. Platt}
\author[mymainaddress]{K. Prasad}
\author[mymainaddress,mythirdaddress]{J. Sander}
\author[mymainaddress]{S. Upadhyayula}

\address[mymainaddress]{Department of Physics \& Astronomy, Texas A\&M University, College Station, TX 77843, USA}
\address[mythirdaddress]{Department of Physics, University of South Dakota, Vermillion, SD 57069, USA}

\begin{abstract}

A dedicated facility has been commissioned for Cryogenic Dark Matter Search (CDMS) detector fabrication at Texas A\&M University (TAMU). The fabrication process has been carefully tuned using this facility and its equipment. Production of successfully tested detectors has been demonstrated. Significant improvements in detector performance have been made using new fabrication methods/equipment and tuning of process parameters.

\end{abstract}

\end{frontmatter}


\section{Introduction}
The quest for an understanding of Dark Matter is at the forefront of particle physics today. One proposed constituent of Dark Matter is a particle called the WIMP (Weakly Interacting Massive Particle). Many experiments around the world are attempting to directly detect this particle in order to further understand the nature of Dark Matter. CDMS (Cryogenic Dark Matter Search) is one of these experiments, utilizing ground based detectors sensitive to nuclear recoils caused by direct collisions with these particles. These detectors, composed of germanium or silicon, are instrumented to detect deposition of both ionization and phonon energy upon interaction with an incident particle. Using these two signals and clever detector design\cite{akerib}, electron recoils (characteristic of the dominant background signals) can be discriminated from nuclear recoils (characteristic of a WIMP-like interaction). To continue probing new parameter space, greater sensitivity is needed for each generation of experiment, requiring significant improvements in the detector technology used to reject electron recoils and fabrication throughput. While many of these improvements are being made in readout electronics and analysis procedures, the underlying quality of the initial data, and therefore the quality of the detectors, is a fundamental limit\cite{cdms2010}. Detector uniformity and consistency are also important, particularly when scaling up in size and quantity of detectors. Great strides in both of these areas have been made at the TAMU facilities. Serving as a second fabrication facility and now the dominant polishing facility, the increased throughput will allow the next generation experiment to produce the required detector payload on a competitive timescale.

\section{Fabrication}
\label{fab}
\begin{figure}[ht]
    \centering
    \includegraphics[width=.95\textwidth]{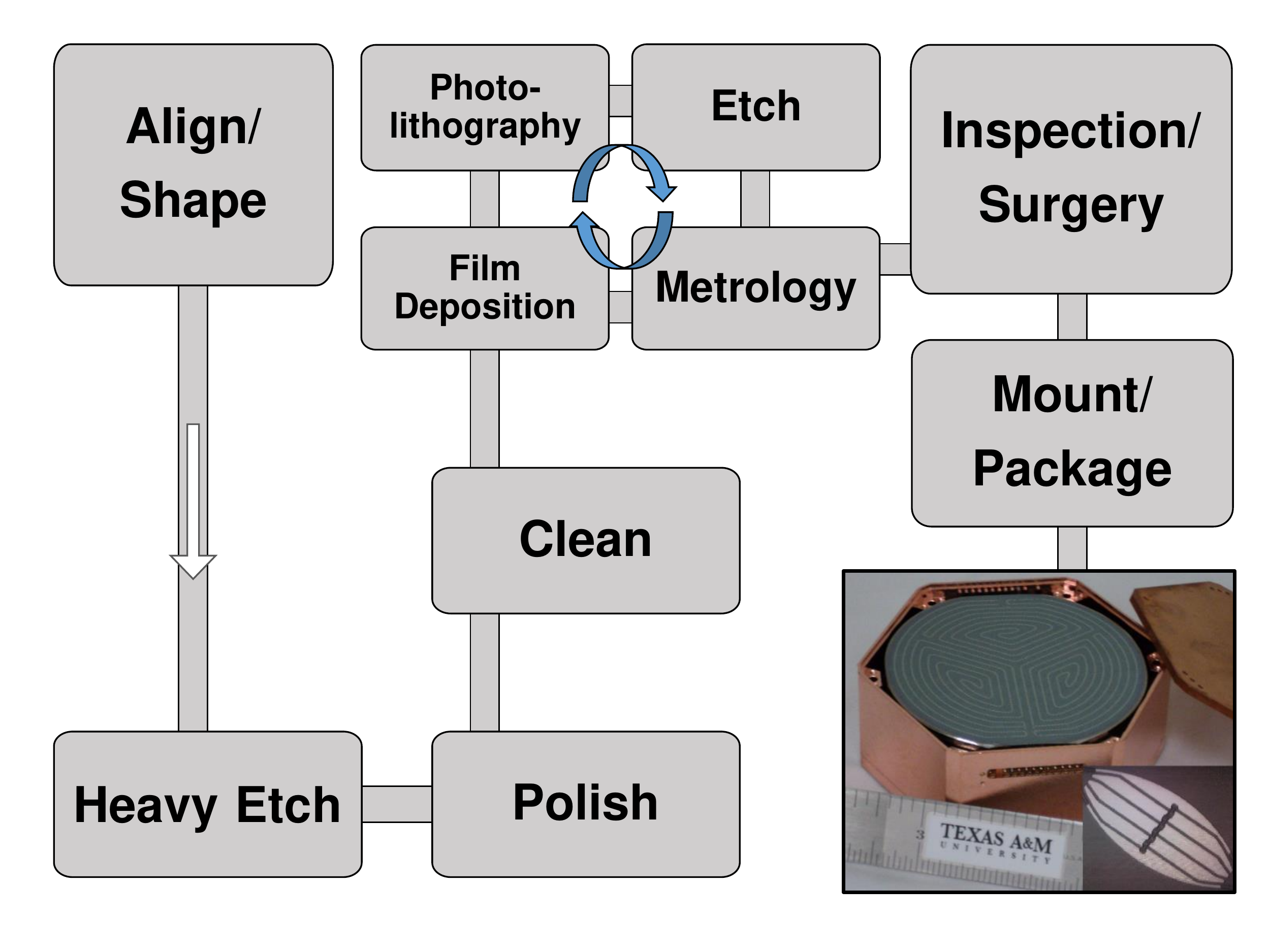}
    \caption{Process flow chart from raw substrate to completion (see Section \ref{fab}).}
    \label{fabflow}
\end{figure}

CDMS detectors are produced using techniques and equipment similar to those in typical semiconductor fabrication processes. The process begins with high quality semiconductor substrates (germanium or silicon) and uses photolithography to etch deposited films into circuit structures (see Figure \ref{fabflow}). One significant difference, however, is that the substrates used for CDMS detectors are much thicker, ranging from 10mm in original designs to 33.3mm in current production. For this reason, semiconductor equipment and processing typically used for $\sim$1mm thick substrates have been modified and tuned for these larger detector geometries.
 
\subsection{Substrate Materials}
Detectors are fabricated on high purity germanium and silicon substrates. For detector quality germanium substrates, ``high purity'' equates to impurity levels typically on the order of 10$^{10}$cm$^{-3}$ . These are grown using the Czochralski Process. Substrates used for this experiment typically have dislocation densities of 1000-7000cm$^{-2}$. For silicon detectors, quality is specified/determined by room temperature resistivity. While $>$8 k$\Omega$-cm is the specification for acceptable material, typical detector quality substrates have a resistivity of $>$20 k$\Omega$-cm and are grown using the Float Zone Process. All detectors currently operating in the SuperCDMS Soudan experiment are 76mm diameter x 25mm thick germanium substrates. The next generation will utilize both silicon and germanium detectors, 100mm diameter by 33.3mm thick\cite{sander2013}.
The majority of this R\&D, including establishment and tuning of the fabrication process (specifically film characterization and photolithography steps), is performed using low resistivity commercial ``Prime Grade'' silicon wafers (75mm and 100mm diameter with SEMI Standard\footnote{Standards/specifications available from www.semi.org} thickness and flats). Being much lower in cost, easier to obtain, and easier to clean/prepare than thick substrates, they are a natural choice for practice and R\&D. Once established, fabrication procedures are then tested/confirmed on thick substrates. Low purity/price thick substrates are used for this before fabricating detector quality substrates.

\subsection{Alignment and Shaping}
To improve uniformity and charge collection performance among the detectors, the substrates are shaped and aligned to a specific crystal axis and orientation. Upon delivery from the vendor, the cylindrical substrates are guaranteed to be aligned within $\pm 2\,^{\circ}$ of the target crystal axis, typically $[100]$. For improved ionization drift/collection, they are subsequently re-shaped with the alignment refined to $\pm 0.1\,^{\circ}$. For this reason, the substrates, as purchased, are slightly over-sized in all dimensions to account for material loss in re-shaping. Re-shaping consists of aligning and grinding the substrates' faces, followed by grinding the cylindrical sidewall. A custom fixture has been made to allow the surface plane of the substrate to be manipulated with micrometers to precisely dial in the crystallographic axis to the coordinates of the x-ray diffractometer (XRD) used in this alignment process. For more information on the XRD process, see \cite{xrdbook}. In this setup, a modified Rigaku DMAX-1BX is used with the x-ray source operated at 30kV and 20mA. First, the face of the substrate is positioned/aligned to the point of initial interference with the x-ray beam (which is set to 2$\theta$=$0\,^{\circ}$) with the face parallel to the beam and perpendicular to the goniometer's $\theta$ plane. The goniometer is then set to the Bragg angle of the target crystal axis and a local 2$\theta$ sweep is performed (the width of which is dictated by the alignment tolerance from the vendor). This produces a peak near the Bragg angle which will shift according to aforementioned micrometer adjustments. These adjustments and measurements are made iteratively (gross adjustments at first, followed by fine tuning) until the peak is within the required tolerance of the appropriate Bragg angle. The crystal is then locked into that orientation in the alignment fixture, which is designed such that it can be unmounted from the XRD system and attached to a grinding fixture. This assembly is then placed on a Lapmaster 24C lapping machine (equipped with a 220 grit diamond magnetic plate) such that the substrate feeds into the grinding surface along the crystal axis. After grinding, this surface is measured again using XRD to confirm successful alignment. The second face is then ground parallel to the first using this same fixture. Parallelism of the faces is confirmed using a granite indicator stand. The cylindrical sidewall must then be shaped parallel to the crystal axis. To reduce the chance of chipping during this process, circular plates of glass (1/8'' thick with a diameter 0.25'' larger than the final substrate diameter) are bonded to each face with a wax bonder using $69\,^{\circ}$C quartz wax. The sidewall shape is then defined using a diamond coring fixture. The coring diameter is that of the final substrate specification. The glass plates and quartz wax are then removed. To provide room for interface boards in the detector housings (see Section \ref{inspect}) and ensure all crystals are fabricated in a uniform rotational orientation, flats are ground on the sidewall of the substrate. These are located normal to a specific crystallographic direction ($[011]$ in the case of $[100]$ crystals). To perform this alignment, the crystal is loaded into a custom XRD mount with the previously aligned crystal axis normal to the 2$\theta$ plane and the x-ray beam incident upon the sidewall (with the sidewall now positioned to just slightly interfere with the beam while 2$\theta$=$0\,^{\circ}$). The goniometer is then set to the Bragg angle of the desired flat orientation, and the crystal is rotated about its axis (in the 2$\theta$ plane) using a precision rotary table indexer until the diffracted intensity is maximized (locating the orientation to $\pm 1\,^{\circ}$). Using a custom jig, the two diametrically opposed flats are ground using the Lapmaster 24C. The crystal is then lapped (on the same machine) to its desired thickness.

\subsection{Heavy Etch}

In order to remove substrate surfaces that may have been contaminated by previous processing and/or exposure to radon-containing atmosphere, the substrates are then chemically etched\footnote{This etch recipe is based on a process described in \cite{holmes} modified by Paul Brink and Larry Novak.}. This process removes the outer layer (up to $\sim$250$\mu$m) of material, which is assumed to be contaminated/compromised. Silicon substrate etching has not been used by CDMS in the past, but is currently under development for future detectors. Germanium etching is performed in the following solution:
\begin{enumerate}
\item 3200mL 69\% HNO$_{3}$
\item 640mL 50\% HF
\item 150mL CH$_{3}$CO$_{2}$H (glacial)
\end{enumerate}
The substrate is dipped in the etchant using a modified PTFE wafer cassette (used in all subsequent acid processing) and agitated lightly by hand, followed by a dip in de-ionized (DI) water. This is repeated 10 times. It is then placed in a Verteq 1600-55M spin rinse/dryer for a {\it standard rinse/dry} process (to be referred to as SRD). The SRD process consists of the following steps: 
\begin{enumerate}
\item 35 seconds @600rpm with N$_{2}$ purge and DI spray
\item 230 seconds @1600rpm with heated N$_{2}$ purge
\item 90 seconds @1600rpm with N$_{2}$ purge
\end{enumerate}
Following this step, substrates are stored in nitrogen purged cabinets when not being actively processed, reducing subsequent exposure to ambient radon.

\subsection{Lapping/Polishing}
Photolithographic processing of micron scale features requires a smooth, featureless substrate surface. For this reason, the coarsely lapped, heavy-etched detector faces must be polished. This is accomplished via four sequential steps:
\begin{enumerate}
\item Fine-grit manual lapping
\item Surface shaping polish
\item Scratch-removing polish
\item Final surface treatment polish
\end{enumerate}
Substrates are hand-lapped on a slotted glass lapping plate using 9$\mu$m alumina polishing powder mixed with DI water to form a paste consistency.  This is to remove large features from the surface. The surfaces must then be polished to a specified flatness with a mirror finish, free of visible features (such as scratches or pits) to facilitate uniform film depositions and prevent circuit defects (see Figure \ref{defects}) in subsequent processing.  Polishing is performed on a dual spindle polishing machine. Control of surface curvature (concavity vs. convexity) is maintained with polisher settings and various sizes of polishing pads surfaced with polyester material in a 1:1 mixture of colloidal alumina polishing compound:DI water. This process is carefully controlled such that the final surface has $<$2$\mu$m of total height deviation across the substrate if convex, $<$1$\mu$m if concave (curvature is measured with a desktop laser interferometer). This is to ensure uniform contact with the photo mask (which can conform slightly to convexity but not concavity) during the photolithography process. Small surface scratches resulting from this step are then removed on the same machine using ``regular nap'' polyurethane pads and a fresh mixture of the same polishing slurry. Final surface polishing is performed with ``high nap'' polyurethane pads in a colloidal silica polishing compound. Final surface inspection is performed using a stereo zoom binocular microscope, manually confirming a defect free mirror finish.

\subsection{Cleaning}

Before the polished substrates can be processed into detectors, they must be cleaned carefully. This removes surface contaminants as well as any particulates that may cause defects in subsequent processing (see Section \ref{inspect}). For this reason, the cleaning is performed in a class 100 clean room. Germanium and silicon substrates require different cleaning processes, germanium's being much more time and labor intensive (another benefit of using silicon wafers for R\&D).

\subsubsection{Germanium}
Initial germanium cleaning involves a 5 minute soak in acetone followed by a 5 minute soak in isopropyl alcohol (NOTE: all chemicals used in cleaning and subsequent processing are semiconductor grade). Following a thorough rinse with DI water, the surfaces are manually inspected using a microscope equipped with an LED ring light (especially effective for identifying particles on the surface as it exposes diffuse features). If particulate count is unacceptable ($\ge$10cm$^{-2}$), the previous chemical process is repeated, and the crystal is dried using a filtered nitrogen gun. If particulate count is still unacceptable, the substrate is rinsed with methanol and manually wiped with a PVA cleaning brush. If the surface condition is still unacceptable, the methanol and brush wipe is repeated as necessary. Otherwise, the substrate proceeds to the oxide removal step. For this, a mixture of 3:1 DI water:50\%HF is prepared in which the substrate is submersed for 5 minutes followed by a 3 minute soak in DI water. This is repeated three times and followed by surface inspection. If particulate count has become unacceptable, a methanol rinse and brush wipe are repeated as necessary. Upon completion, the substrate is placed in the oven at $120\,^{\circ}$C for 10 minutes to bake out remaining moisture.

\subsubsection{Silicon}
Silicon substrates also receive chemical cleaning, but have not shown the need for manual particulate removal. In the cleaning process, the substrates are initially doused with methanol then isopropyl alcohol, followed by SRD. To remove metals and organic contaminants, a Piranha clean process is used. This consists of a 20 minute dunk in the following solution (heated to 55$^\circ$C):
\begin{enumerate}
\item{600mL 30\% H$_2$O$_2$}
\item{700mL 98\% H$_2$SO$_4$}
\item{1050mL 0.250N H$_2$SO$_4$}
\end{enumerate}
Substrates are then soaked in 55$^\circ$C DI for 1 minute, manually agitated once every 15 seconds. To remove the native oxide from the substrates' surfaces, they are then placed in the following solution for 20 seconds:
\begin{enumerate}
\item{2700mL DI}
\item{50mL 50\% HF}
\end{enumerate}
The substrates are then dipped again in 55$^\circ$C DI for 1 minute, manually agitated once every 15 seconds. To remove ionic and heavy metal atomic contaminants from the substrates' surfaces, the substrates are submersed in the following solution, heated to 70$^\circ$C, for 15 minutes:
\begin{enumerate}
\item{1750mL DI}
\item{325mL 30\% H$_2$O$_2$}
\item{300mL 37\% HCl}
\end{enumerate}
This is followed by SRD and a 5 minute dehydrate in the oven at 120$^\circ$C. The cleaning process seals the substrate with a thin oxide layer which is removed in the sputtering system prior to film deposition (see Section \ref{segi}).

\subsection{Thin Film Deposition/Tuning}
\label{segi}

\begin{figure}[ht]
    \centering
    \includegraphics[width=.95\textwidth]{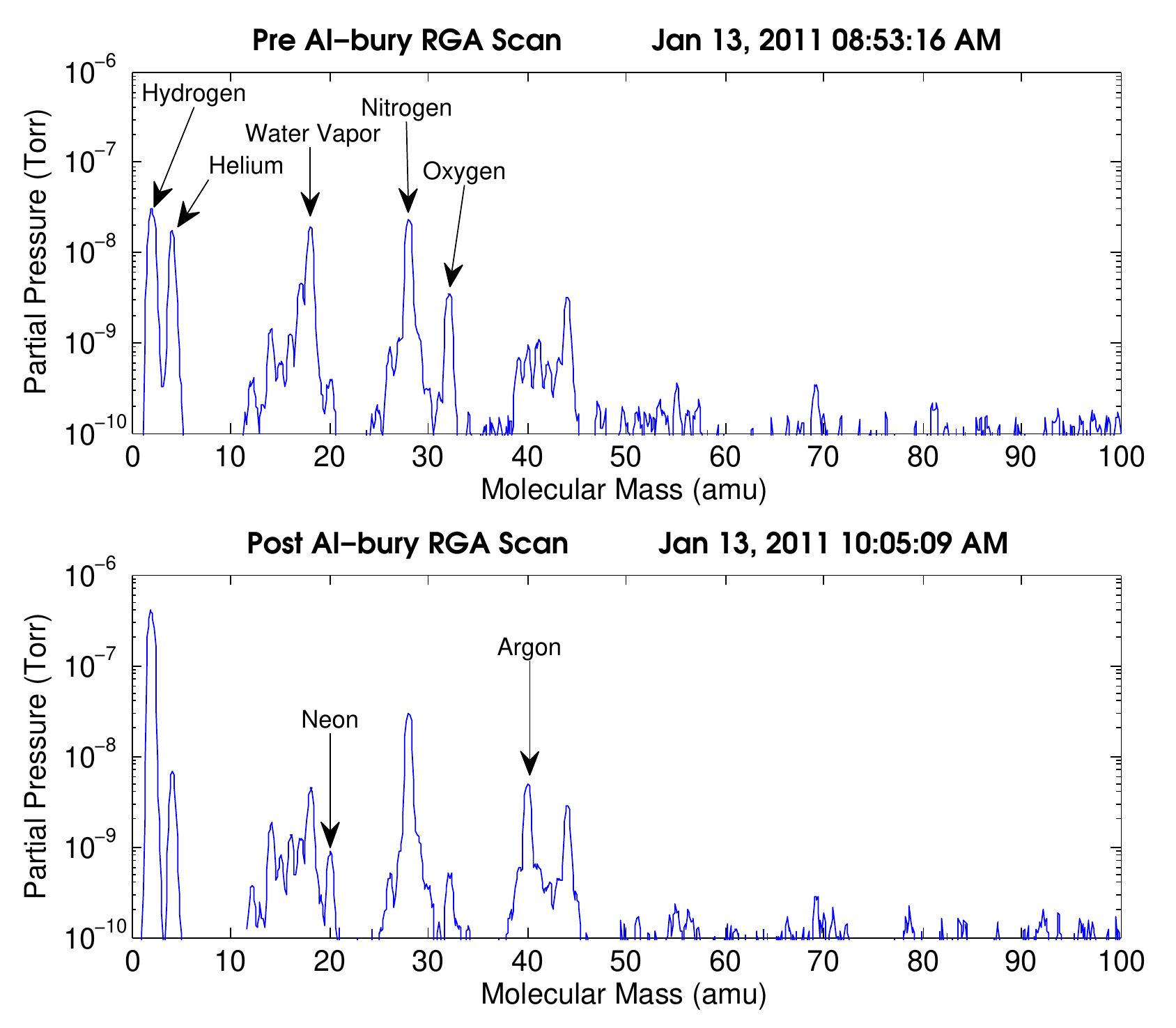}
    \caption{Example RGA spectra of partial pressures in the SEGI process chamber at various atomic masses before ({\it top}) and after ({\it bottom}) aluminum getter deposition (see Section \ref{segi}), demonstrating the efficacy of this process in reducing oxygen and water vapor levels.}
    \label{rga}
\end{figure}

The films that form the final circuit and sensors of the detector are deposited using a customized plasma sputtering deposition system. Precise and repeatable process control is vital in the deposition of these films as they dictate the quality of the final circuit features (see Sections \ref{rrr}, \ref{tctune}, and \ref{tes}). The deposition system used in this process is a Perkin Elmer 4400 Delta with PLC/PC interface automated by Semiconductor Engineering Group, Inc. (SEGI), which has been modified for this fabrication process. It can simultaneously sputter 8 substrates, 100mm in diameter (or 6, 150mm diameter) and is composed of two main vacuum chambers: the load lock and the process chamber, separated by a gate valve. The load lock is the location in which substrates are initially loaded, which is then pumped to 4.0x10$^{-6}$ Torr using a turbo-molecular pump (added for this process). This allows the process chamber to stay isolated from atmospheric contaminants at all times (specifically when the substrate is transferred through the gate valve into this chamber). To further reduce contaminants, a pre-coat of Aluminum is sputtered in the process chamber as a getter. This process removes traces of O$_{2}$ and H$_{2}$O (see Figure \ref{rga}) as well as other contaminants which can alter film characteristics, further improving process stability. The substrates are then transferred into the process chamber, which is subsequently pumped back to a base pressure of 9.0x10$^{-7}$ Torr. The process chamber is cylindrical ($\sim$1m in diameter) and can simultaneously accommodate three different targets of sputtering materials. The chamber is equipped with aluminum, tungsten, and silicon targets, all of which are 99.999\% pure. After substrates are loaded in the load lock, the rest of the deposition process is entirely automated, including everything from rotation and height settings of the table on which the substrates sit to setting/maintaining the chamber and plasma conditions. Table height settings are customized for each substrate thicknesses to maintain a constant target-substrate distance. Other settings of particular note are the voltage and power supplied to the target (DC or RF), DC bias applied to substrates during deposition, flow of Argon into the chamber, and optional RF pre-etch.

The process chamber is equipped with an RGA (Residual Gas Analyzer) which measures the contents of the gas in the process chamber. The RGA can be used in two modes: plotting the entire spectrum at once, showing peaks at various masses corresponding to contaminants in the chamber (see Figure \ref{rga}), or plotting the levels of a chosen contaminant over time. A high capacitance valve is implemented to allow the RGA to operate at high vacuum levels as well as in-process levels ($\sim$10mTorr). This allows in-situ analysis of any possible gas contaminants during the deposition process.

\begin{figure}[ht]
    \centering
    \includegraphics[width=1\textwidth]{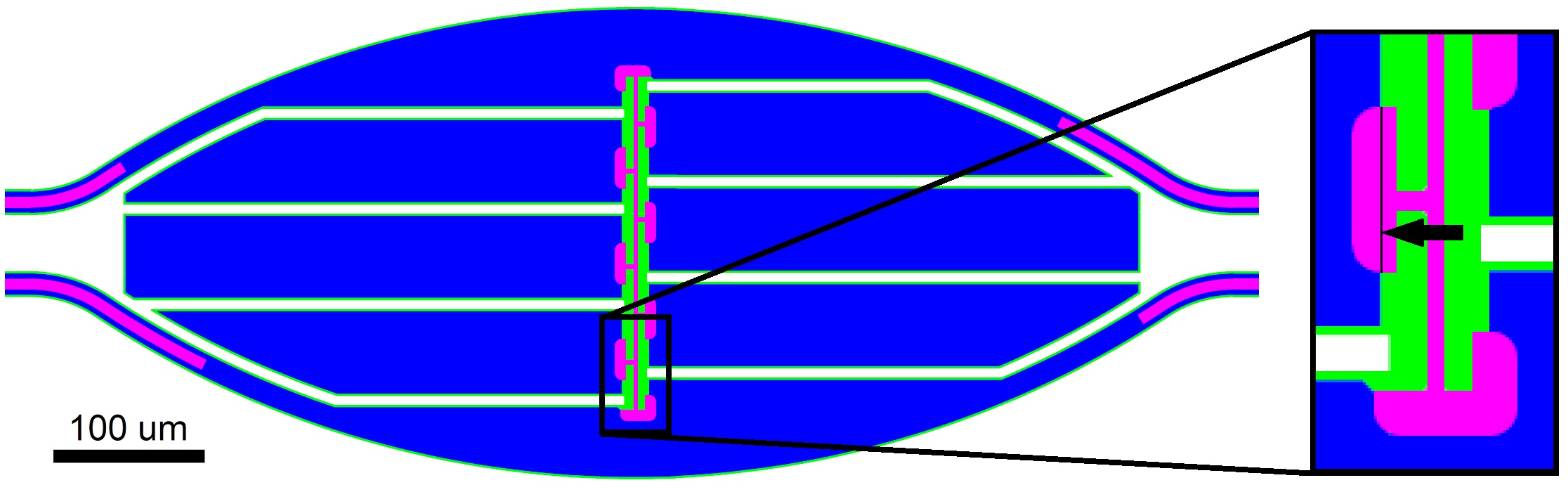}
    \caption{Image of a single phonon sensor and magnified inset of TES line and ``waterfall'' region (arrow indicates waterfall boundary, see Figure \ref{wfallview} for closeup and Section \ref{tes} for description). Each color corresponds to the exposed film on the final structure:  Blue=Aluminum (mask $\#$1, trilayer mask), Pink=Tungsten (mask $\#$2, TES mask), Green=a-Si (mask $\#$3, ``trench'' mask) (see Section \ref{photo}). The central vertical line is the $\sim$2$\mu$m wide TES, and the large aluminum ``fins'' are the phonon absorbing structures (see Section \ref{rrr}). Note: a-Si underlies all metal features.}
    \label{stzm}
\end{figure}

All of the above devices/processes allow the minimization of contaminants and maximization of control and repeatability in the deposition process. Using this system, three thin film layers (designated as the trilayer) are sputtered sequentially on both faces of each substrate: 40nm amorphous silicon (a-Si), 300nm aluminum, and 30nm tungsten. The a-Si layer underlies all final metal circuit structures and is used to protect the substrate surface from aluminum and tungsten etchant chemicals, as well as improving the ionization collection boundary. The aluminum layer forms the phonon collection structures (see Figure \ref{stzm} and Section \ref{rrr}) as well as the circuit lines/electrodes connecting the sensors on the detector. The tungsten layer serves as a cap layer, preventing the aluminum surface from oxidizing (see Section \ref{tes}) and preventing back-sputtering of the aluminum during the subsequent deposition. Before each deposition begins, the target to be used is pre-sputtered for 25 seconds to clean its surface with its shutter closed (to prevent sputtered material from depositing on substrates). Before all depositions, an aluminum shadow mask is manually placed on the substrate face which covers the outer $\sim$1mm of the exposed surface, preventing deposition on this region (see Section \ref{pr}) and the substrate's sidewall surface.
The steps and parameters used for the trilayer deposition are the following: 
\begin{enumerate}
\item 10 minute RF etch, 350 W RF, 50 sccm Ar, 10 mTorr
\item 16 minute a-Si deposition, 500 W RF, 50 sccm Ar, 8 mTorr 
\item 7m18s aluminum deposition, 2.5 kW DC, 40 sccm Ar, 10 mTorr
\item 36 second tungsten deposition, 2.5 kW DC, 40 sccm Ar, 8 mTorr
\end{enumerate}

The films are then patterned photolithographically and chemically etched, forming the majority of the detector circuit (described in Section \ref{photo}). After chemical etching, a 40nm layer of tungsten is sputtered on each face of the substrate. This layer forms the {\it transition edge sensors} (TES's) of the detector (see Sections \ref{tctune} \& \ref{tes}).
 The steps/parameters used for this deposition are as follows:
\begin{enumerate}
\item 10 minute RF etch, 350 W RF, 50 sccm Ar, 10 mTorr
\item 51 second tungsten deposition, 2.5 kW DC, 40 sccm Ar, 8 mTorr, 100V DC bias delivered to substrate
\end{enumerate}
 
\subsection{Photolithography}
\label{photo}

A three step photolithographic process is used to define the circuit features on the substrates. The original process from which this was adapted is described in \cite{brink2009} and \cite{brink2014}. The first step defines the aluminum structures (circuit lines and phonon collecting fins [see Section \ref{rrr}]). The second step defines the tungsten TES features (see Section \ref{tctune}), and the third defines the a-Si structure and substrate trenching regions (see Section \ref{iZip}). The aluminum and tungsten are etched with chemicals, while the a-Si is plasma etched via an RIE (Reactive Ion Etch) process. In all three steps, an etch resistive mask of photoresist is used to protect the features while the exposed films are etched. The photoresist mask pattern is formed via UV transfer/exposure using a master template mask. Chemical processing of the substrates is performed in a class 100 UV-free clean room. 

\begin{figure}[!ht]
    \centering
    \includegraphics[width=.95\textwidth]{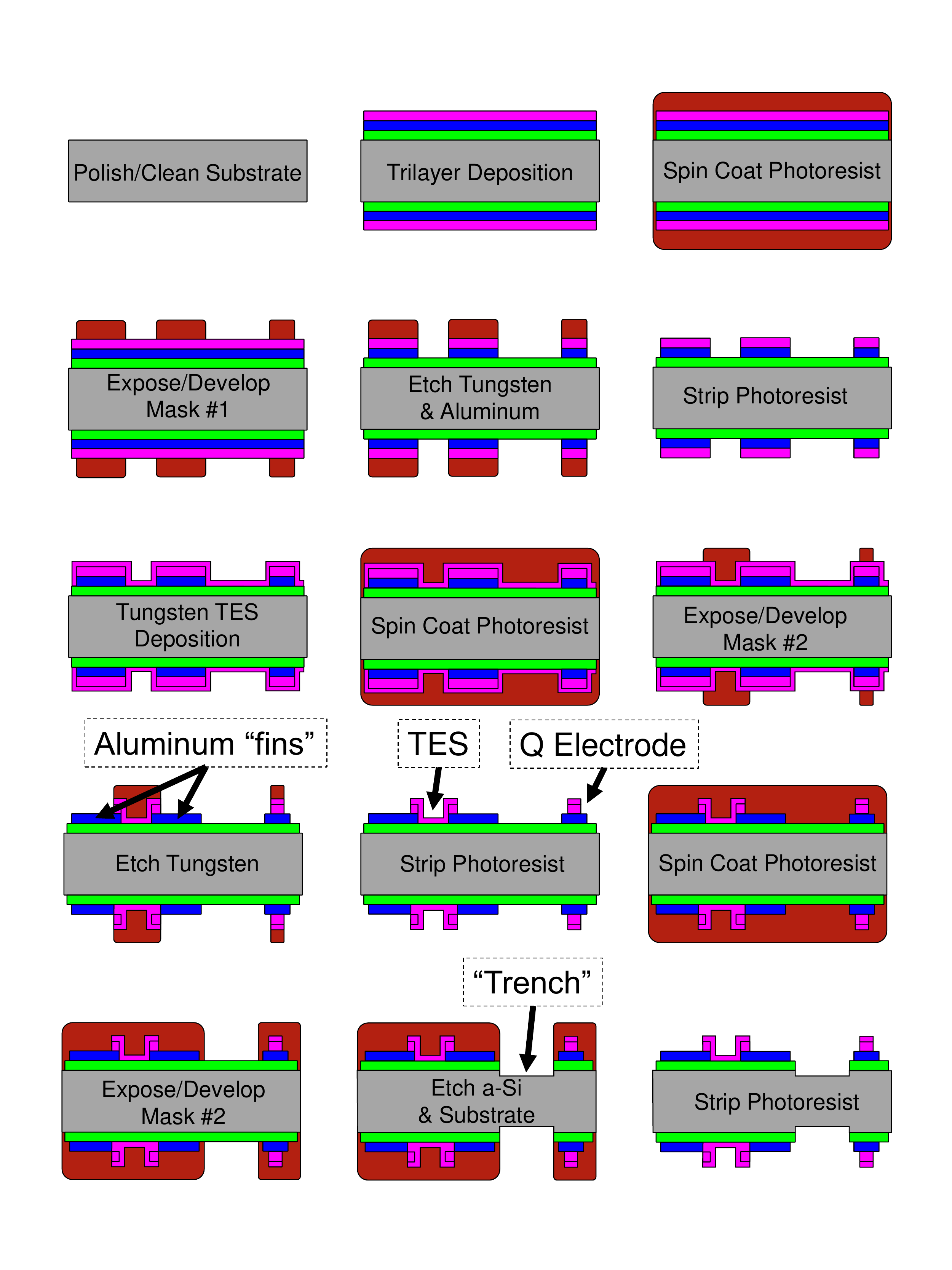}
    \caption{Detector patterning process (not to scale). Individual film layers are shown throughout the deposition/photolithography process (see Section \ref{photo}). Gray=Substrate, Green=a-Si, Blue=Aluminum, Pink=Tungsten, and Red-brown=Photoresist.}
    \label{litho}
\end{figure}

\subsubsection{Trilayer Patterning}
\begin{figure}[ht]
    \centering
    \includegraphics[width=1\textwidth]{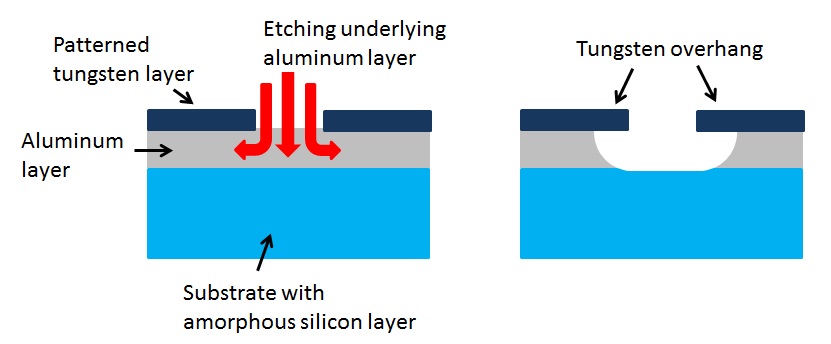}
    \caption{Diagram depicting the tungsten overhang issue caused by the isotropic aluminum etch process (not to scale). Figure from \cite{kunj}.}
    \label{overhang}
\end{figure}

After the trilayer deposition (see Section \ref{segi}), a Solitec 5110-SJ spin coater is used to spin coat Shipley Microposit S1811 photoresist on both faces of the substrate. To create and maintain a vacuum seal between the substrate and spin coater's chuck, 0.032'' thick elastomer “skirts” are stretched around the substrate sidewall. These are removed and discarded after the spin coat process (see Section \ref{pr} for more information on the photoresist layer/process). The substrate is then placed in the oven (in a PTFE cassette, covered with aluminum foil to protect the fresh photoresist from particulates) at $120\,^{\circ}$C to soft bake the photoresist (see Table \ref{bakes} for bake times). After baking, the crystal is allowed to cool to room temperature. The cooling process is accelerated with a gentle stream of filtered nitrogen gas on each face. The substrate is then ready for the photolithographic mask transfer/exposure process. An OAI 206-094735 contact aligner with a 350W Hg g-line UV lamp is used to expose each face for 5.3s at 8.15mW/cm$^2$ using mask $\#1$, the trilayer mask (see Figure \ref{stzm}). Special care must be taken not to scratch the backside photoresist layer when placing the substrate on the stage (and when flipping the substrate for exposure of the second face). The UV intensity is confirmed before each exposure using an OAI 0308 UV meter tuned to 436nm (g-line). The pattern is then developed using Shipley Microposit MF-319 developer, mildly agitated by hand, until completion. This is judged by eye, typically taking 70-90 seconds. The substrate is gently rinsed in DI water after development, then proceeds to SRD. The photoresist is then inspected to confirm successful development (robust replication of mask structure). The substrates are then returned to the oven at $120\,^{\circ}$C to hard bake the photoresist (see Table \ref{bakes} for bake times). They are then cooled to room temperature with the assistance of gentle nitrogen gas flow. The tungsten layer is etched for 6m30s using 30\% H$_{2}$O$_{2}$ with 2 gentle manual agitations at 1 minute intervals, followed by SRD. The aluminum layer is etched using Cyantek Al-11. This typically consists of 5-6 iterations of the following: 45s Al-11 dunk with constant gentle agitation, followed by a 15s DI rinse. The aluminum etch leaves a slight overhang of the tungsten cap layer, due to the isotropic nature of the reaction (see Figure \ref{overhang}). Intermittent DI rinses are used to control the temperature of the exothermic etch reaction\cite{williams}, improving etch uniformity and reducing the undercut/overhang issue. When all exposed aluminum appears to have vanished, the substrate receives an additional 15s of Al-11 etch to ensure no aluminum remains, then proceeds to SRD. To remove the overhang feature, another tungsten etch is performed (see Section \ref{tes}). This consists of a 10 minute submersion in 30\% H$_{2}$O$_{2}$, with 2 gentle agitations every 2 minutes, followed by SRD. At this point, the circuit pattern is carefully inspected to confirm successful etching and preserved photoresist integrity. The photoresist layer is removed using a 20 minute dip in Shipley PRX-127 at $45\,^{\circ}$C, with 2 gentle agitations every 5 minutes, followed by SRD. As a final cleaning precaution, the substrate is submerged in Baker PRS-1000 for 10 minutes at $45\,^{\circ}$C, followed by SRD. Etched features are then inspected (and again after each subsequent photolithography/etch cycle), monitoring for defects and critical circuit feature dimensions. The substrate is then placed in the SEGI under vacuum overnight to boil off any moisture before the following tungsten (TES layer) deposition.

\subsubsection{TES Patterning}
The second deposition, that which forms the TES tungsten layer, is then performed (see Section \ref{segi}). After this deposition, the substrate receives the same spin coat, soft bake, alignment, exposure, develop and hard bake process as previously mentioned. The mask used for this layer (mask $\#2$, see Figure \ref{stzm}) defines the TES structures on the circuit. After hard bake and cooling, the tungsten is etched in 30\% H$_{2}$O$_{2}$ for 12 minutes, with 2 gentle agitations every 2 minutes, followed by SRD. It should be noted that this step etches all tungsten not covered by the mask, including the tungsten cap layer from the mask $\#1$ structures. Therefore, anywhere that masks \#1\&2 coincide, all four film layers remain. Otherwise, mask $\#$2 defines structures with only TES tungsten on top of a-Si (see Figure \ref{litho} and Section \ref{tes}). The photoresist is then inspected for integrity and removed with the same PRX-127 and PRS-1000 process as before (aside from PRX-127 time reduction to 15 minutes). The substrate is then placed under vacuum overnight to remove moisture (improving adhesion of subsequent photoresist coat). 

\subsubsection{a-Si Patterning and ``Trenching''}
The last photolithography step defines the a-Si structure with mask $\#3$, using the same spin coat, soft bake, alignment, exposure, develop, hard bake, and cooling process as previous steps. After hard bake and cooling, the a-Si is etched in a modified Tegal 903C reactive ion etcher using 8 iterations of the following etch and cool down steps: 
\begin{enumerate}
\item 18 second etch, 400 W RF @ 13.56 MHz, 18 sccm SF6, 50 sccm He, 1100 mTorr 
\item 7 minute purge/cool down, 50  sccm He, 900 mTorr (limited by Helium MFC)
\end{enumerate}
See Section \ref{iZip} for more information on this etch.
 The photoresist is then removed with the same PRX-127 and PRS-1000 process as before (with original PRX-127 time of 20 minutes).

\begin{table}[ht]
\centering
\begin{tabular}{|c||c|c|}
\hline
Size (Dia. x Thickness) & Soft Bake& Hard Bake\\
\hline
76mm x 10mm & 20m & 1h30m\\
76mm x 25mm & 25m & 2h\\
100mm x 33.3mm & 28m & 2h20m\\
\hline
\end{tabular}
\caption{Bake times for various substrate sizes. Thin (practice) wafers soft bake for 1m50s on a $115\,^{\circ}$C hot plate, and hard bake for 15 minutes in the oven at $120\,^{\circ}$C.}
\label{bakes}
\end{table}

\subsection{Inspection/Surgery/Mounting}
\label{inspect}
\begin{figure}[ht]
    \centering
    \includegraphics[width=1\textwidth]{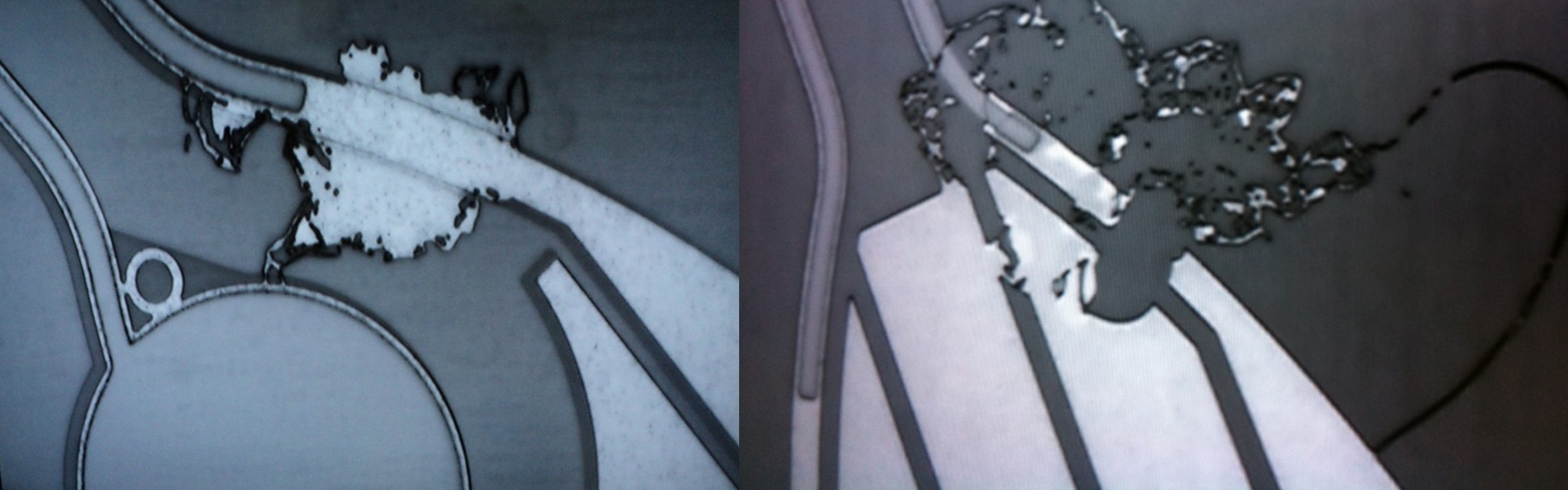}
    \caption{{\it Left)} Example of an un-etched section of Aluminum, causing a possible short. {\it Right)} Example of photoresist failure allowing unwanted aluminum etching, breaking circuit continuity, and requiring a ``surgery'' wire bond. Both examples were likely caused by particulates on the detector surface during early processing.}
    \label{defects}
\end{figure}

It is possible for defects to arise in the photolithography process which can prevent a detector from operating as desired. For this reason, every element of every detector circuit is manually inspected using a microscope. This step is crucial to successful detector fabrication as micron scale defects can knock out an entire sensor channel. Defects of concern include areas of missing metal, causing breaks in the circuit continuity as well as metal films that did not etch properly, causing shorts (see Figure \ref{defects}). In the case of open circuits due to breaks in metal continuity, a Kulicke \& Soffa 4523AD wire bonder (with a DewyL Tool MCSOE-1/16-750-45-C-2025-M wedge and 0.00125'' diameter 99\% Al 1\% Si wire) is used to connect the isolated metal regions with wire bonds. The circuit is designed with extra metal pads (bonding locations) to make this task easier. In the case of a defect causing a short circuit, repairs can be made using one of two options: 
\begin{enumerate}
\item Manually abrading the film with the wedge of the wire bonder to eliminate the unwanted electrical connection (an auxiliary wedge should be used for this, preventing damage to the bonding wedge) 
\item Using a localized droplet of the proper chemical etchant to remove the metal. This can also be used during the photolithography process (depending on the nature/location of the defect), utilizing the protective photoresist mask, thus minimizing chemical damage to nearby structures. 
\end{enumerate}

The detectors are then mounted in hexagonal OFHC copper housings (see Figure \ref{detector}) which include Detector Interface Boards (DIBs). The detector electrodes have large bond pads near the DIBs, used to wire bond the detector channels to copper traces on the DIBs. This provides a feed-through to the outside of the grounded housing as well as a rigid connection to external hardware. 

\section{Process Tuning, Results, and Improvements}

\subsection{Aluminum Film}
\label{rrr}

To efficiently read out phonon energy, the phonons are first absorbed in the superconducting aluminum ``fins'' (see Figure \ref{stzm}) where the energy is used to dissociate Cooper pairs which split into pairs of quasi-particles. These quasi-particles must diffuse through the aluminum to reach the tungsten TES where they are essentially trapped due to the tungsten's lower superconducting bandgap value (see Section \ref{tes}). A crucial property of the aluminum is a high quasi-particle diffusion length. This length is inhibited by impurities found in the aluminum.

Quasi-particle diffusion length is important to monitor and control. To quantify the quality of the aluminum film, a measurement of the {\it residual resistivity ratio} (RRR) is performed. This is the ratio of the film's resistivity at room temperature to its resistivity at 4K. A higher RRR value corresponds to a higher quality film (i.e. one with fewer impurities) \cite{RRR}. A RRR value of $\>$10 has been deemed sufficient for a well functioning device in these detectors, but films have been tuned using the SEGI to routinely achieve a value of $\sim$16. 

\subsection{T$_{c}$ Tuning}
\label{tctune}
\begin{figure}[ht]
    \centering
    \includegraphics[width=1\textwidth]{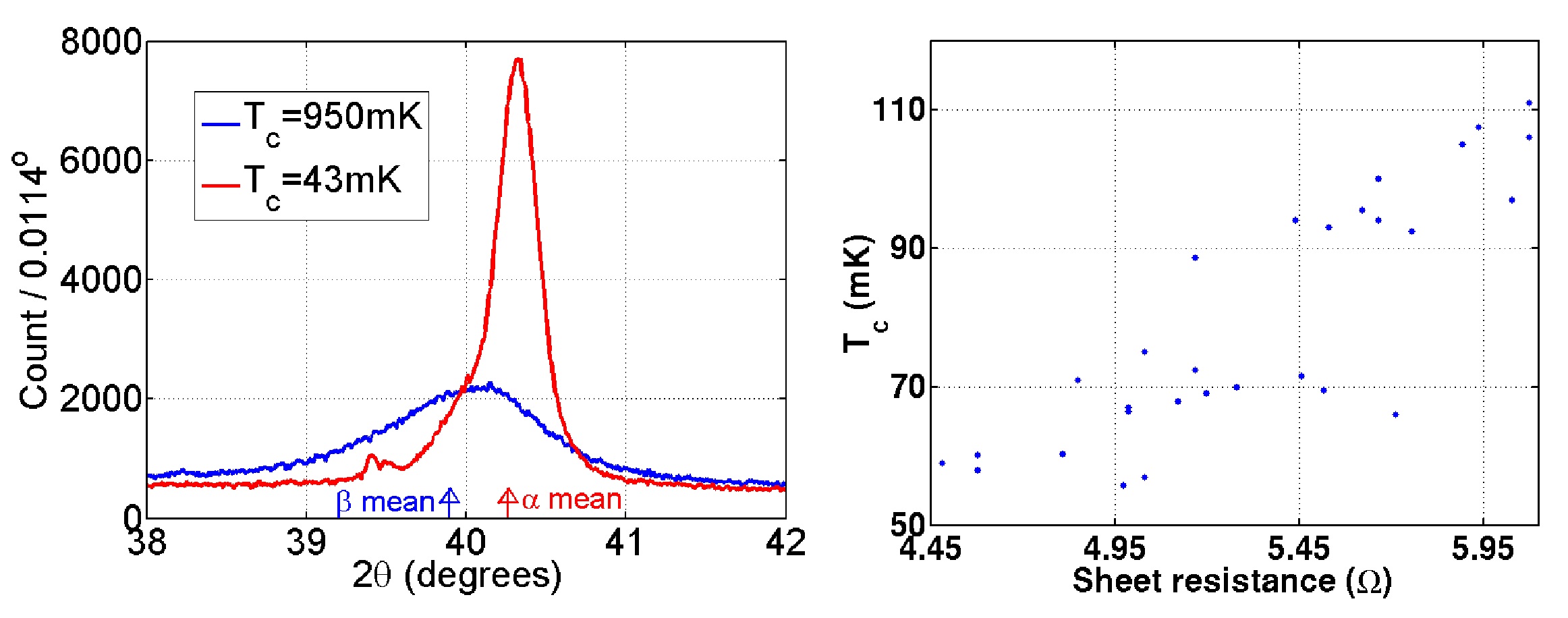}
    \caption{{\it Left)} XRD spectra showing discrimination between samples of differing $\alpha$:$\beta$ concentrations and their resulting T$_{c}$'s. Also marked are the locations of the peaks of pure $\alpha$ and $\beta$ phase films. An algorithm fitting two Gaussian functions (centered at these values) provides an estimate of phase ratios, and therefore T$_{c}$'s, of the films. {\it Right)} Plot showing correlation of critical temperature vs. sheet resistance of similar thickness films (40$\pm$4nm) \cite{kunj}. These room temperature characterization methods allow film deposition parameters to be tuned without the time or monetary expense of dilution refrigerators (see Section \ref{tctune}).}
    \label{xrdtvr}
\end{figure}

Optimal detector readout relies on the TES sensors being held at specific temperatures in their superconducting-transition resistance curves. The second tungsten deposition forms the TES layer. Consequently, this deposition must be carefully tuned to produce tungsten of a uniform, consistent, and precisely-controlled critical temperature (T$_{c}$). Critical temperatures of thin tungsten films are largely dictated by the ratio of $\alpha$ to $\beta$ phase in the material. This is due to the fact that the $\alpha$-W exhibits a T$_{c}$ of 15 mK \cite{lass} while $\beta$-W can have T$_c$'s ranging from 1 to 4 K \cite{lita}. Utilizing this and the fact that the two phases have different crystallographic structures (and therefore, different Bragg angles), provides a technique of roughly estimating the T$_{c}$ of a given sample at room temperature using XRD\cite{lita} (see Figure \ref{xrdtvr}). This technique is useful for tuning film samples to have high $\alpha$:$\beta$ ratios (T$_c$'s closer to the desired range), but in this range, the ratio becomes so heavily $\alpha$-dominated that differences in $\beta$ concentrations become indistinguishable, making T$_c$ predictions difficult. To finely tune deposition parameters to the $\sim$80mK target, a dilution refrigerator is used to physically measure the resistance transition as the sample is cooled past its T$_{c}$ and again as it warms up. With this feedback, depositions with different sputtering power, substrate bias, and argon pressure were produced and tested, creating films of varying T$_{c}$'s. In this process, a correlation was established connecting room temperature resistivity of the films to their T$_{c}$ (see Figure \ref{xrdtvr}, {\it right}), allowing recipes to be roughly tuned and chosen with simple room temperature measurements (sheet resistance measured with a 4-point probe, corrected for film thickness to calculate resistivity)\cite{lita}. Using these processes, a recipe was chosen to produce films possessing the desired T$_{c}$. Current experimentation with devices of varying T$_{c}$'s rely heavily on the resistivity-T$_{c}$ correlation, saving significant time and money required for dilution refrigerator tests.

\begin{figure}[ht]
    \centering
    \includegraphics[width=1\textwidth]{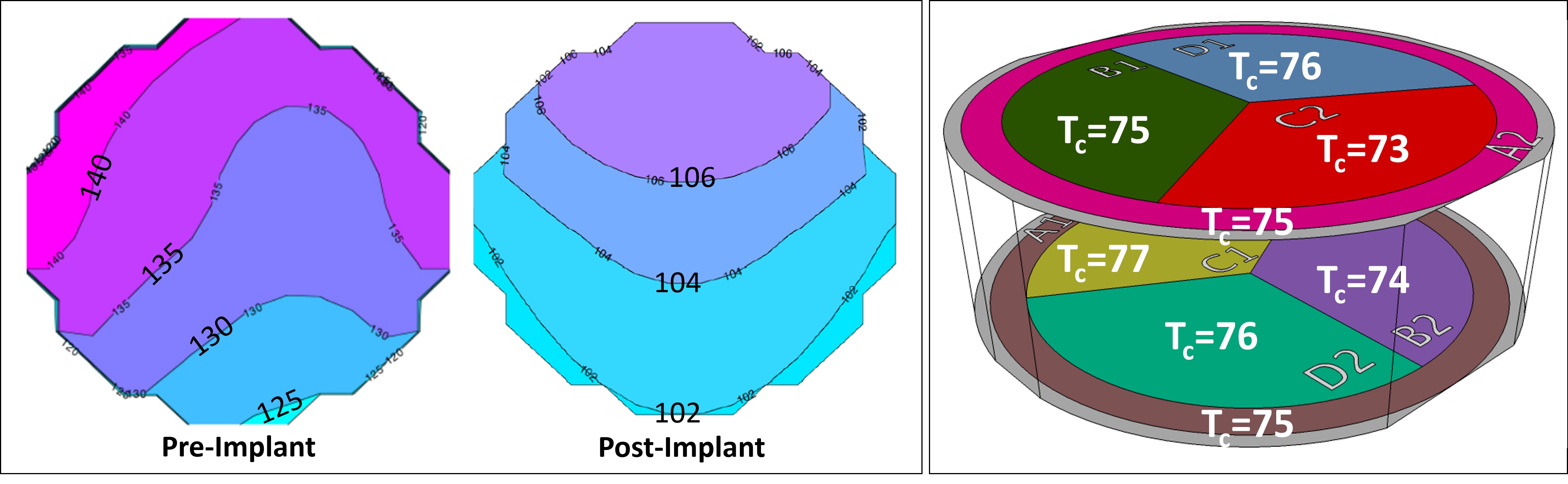}
    \caption{{\it Left)} Example of W T$_c$ variation (in mK) across a sample substrate face from the previous CDMS deposition system. {\it Middle)} T$_c$ variation of same film {\it after} ion implant compensation (see Section \ref{tctune}). Figure adapted from \cite{tcmapnote}. {\it Right)} T$_c$ measurements (see Table \ref{g9data}) from the 8 phonon channels of G9F, a detector fabricated at TAMU, demonstrating T$_c$ uniformity {\it without} ion implantation. All T$_c$'s are in mK.}
    \label{tcmap}
\end{figure}

In addition to depositing films with carefully tuned, repeatable T$_c$'s, the SEGI has demonstrated the ability to produce films with much higher T$_c$ uniformity across the substrate surface. Previous systems have had large T$_c$ gradients across the face of detectors, beyond an acceptable limit (see Figure \ref{tcmap}). To correct this issue, T$_c$ distributions must first be mapped (requiring detector testing in a dilution refrigerator), followed by ion implantation of $^{56}$Fe (specifically into the TES's) to correct for the measured T$_c$ gradient, a process described in \cite{implant}. Films deposited in the SEGI, however, have demonstrated uniformities as good or better than typical {\it post-implant} samples from other systems. This ``as-delivered'' uniformity circumvents a full round of millikelvin testing (T$_c$ mapping) {\it and} ion implantation, increasing throughput rates. The consistency and uniformity of films produced by the SEGI may allow the test process to largely avoid T$_c$ testing, aside from periodic verification. Circuit continuity tests can be accomplished at higher temperatures (up to $\sim$1K), meaning these detectors may be able to avoid dilution refrigerator testing as a whole during high throughput periods. With improved production throughput rates, the bottleneck is shifted from fabrication to testing, exaggerating the importance of these consistency and uniformity improvements.

\begin{table}[ht]
\centering
\begin{tabular}{|c||c|c|}
\hline
Channel & T$_c$ (mK) & R$_n$ ($\Omega)$\\
\hline
A1 & 75 & 0.64\\
B1 & 75 & 0.59\\
C1 & 77 & 0.59\\
D1 & 76 & 0.59\\
A2 & 75 & 0.69\\
B2 & 74 & 0.56\\
C2 & 73 & 0.54\\
D2 & 76 & 0.56\\
\hline
\end{tabular}
\caption{Critical temperatures and ``normal'' resistance values (R$_n$) for the 8 phonon channels of detector G9F (see Section \ref{tctune}). R$_n$ is the resistance of the channel while the aluminum is superconducting, but the tungsten is normal (held at a temperature significantly above its T$_c$).  Note: Channels A1 and A2 are outer channels (see Figure \ref{tcmap}) and have higher R$_n$ values due to their sensor layout.}
\label{g9data}
\end{table}

\subsection{Conformal Film Deposition}
\label{tes}

Controlling the fabrication quality of the aluminum-tungsten interface (to maximize quasi-particle diffusion into the TES) is important because phonons absorbed in the aluminum only contribute to the measured phonon signal if they are able to drift into the tungsten. When quasi-particles drift from the aluminum ``fins'' to the overlapping TES structure (see Figure \ref{stzm}), they must first drift into the intermediate tungsten cap layer. Since this intermediate tungsten layer is deposited immediately after the aluminum layer {\it without breaking vacuum}, no oxide is able to form between the two. Without this cap layer, an oxide forms on the aluminum surface before the TES layer deposition and inhibits the diffusion of quasi-particles from one film to the other. While the cap layer does oxidize slightly, the oxidation is easily removed with the RF etch which precedes the TES film deposition, forming a more favorable interface between the two tungsten layers. 

The quasi-particle propagation from the ``fins'' to the TES's is aided by the bandgap disparity arising from the aluminum and tungsten films' contrasting T$_{c}$'s (aluminum's T$_{c}$ of $\sim$1.2K equates to a gap energy of 0.18meV compared to the tungsten's gap energy of $\sim25\mu$eV)\cite{kunj}. Due to the magnitude of disparity in band gap energies, a process of quasi-particle multiplication can even occur at this boundary\cite{quasi}. 

\begin{figure}[ht]
    \centering
    \includegraphics[width=1\textwidth]{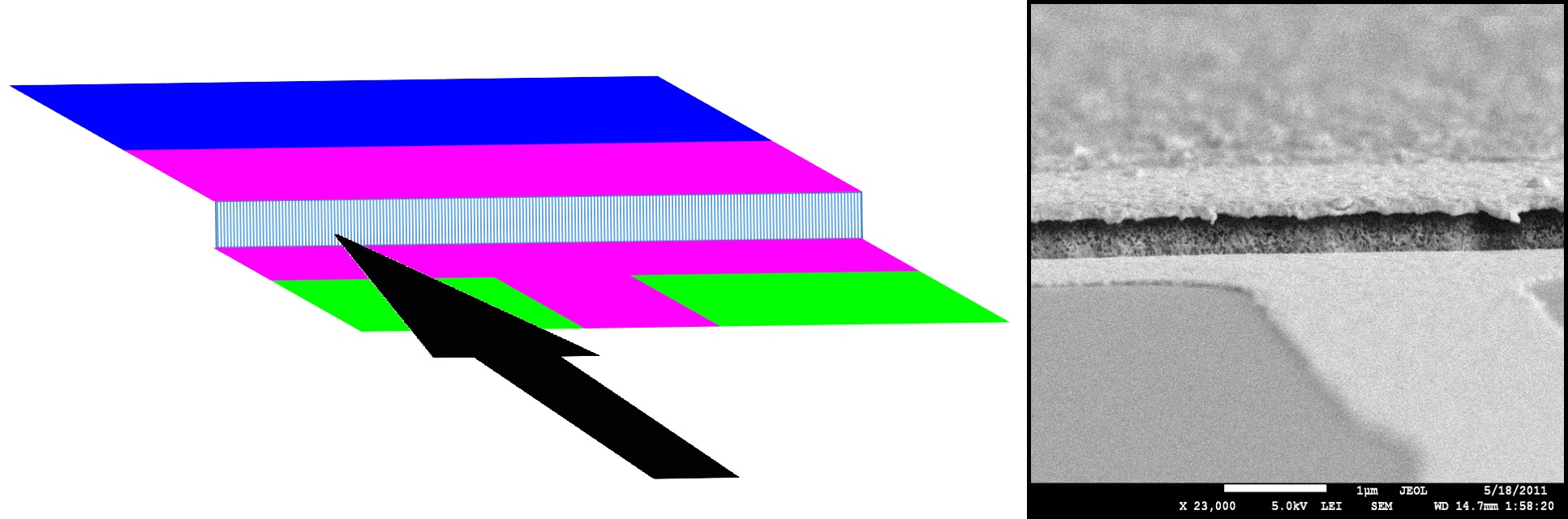}
    \caption{Close-up and SEM image of ``waterfall'' boundary (see Section \ref{tes}). Location on phonon sensor and perspective are indicated by the arrow, referencing Figure \ref{stzm}.}
    \label{wfallview}
\end{figure}

\begin{figure}[ht]
    \centering
    \includegraphics[width=1\textwidth]{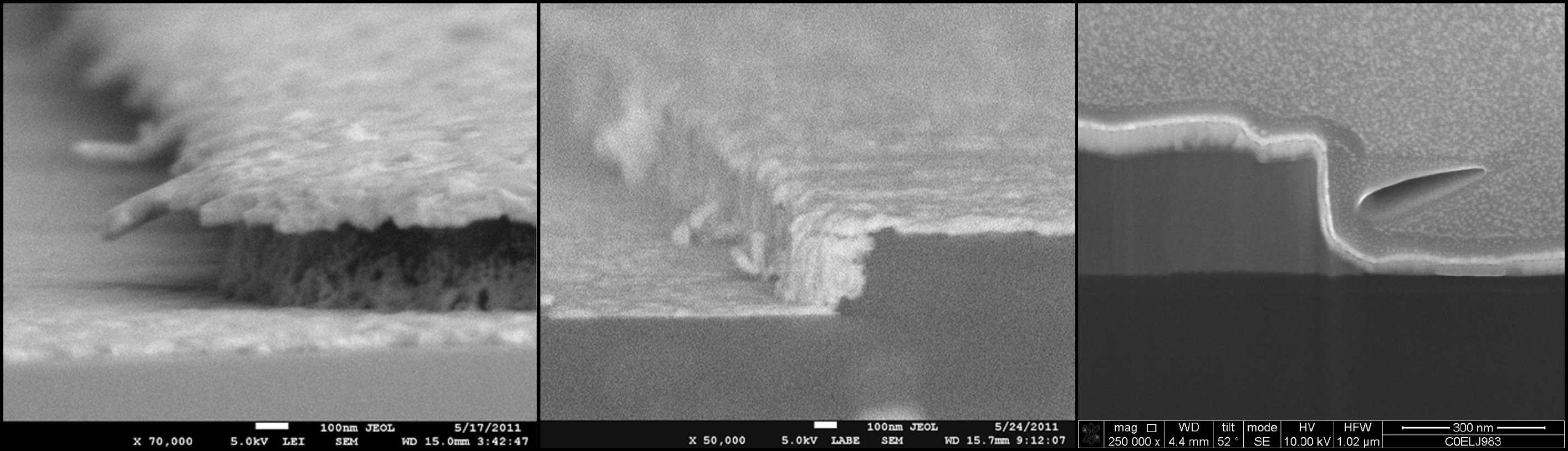}
    \caption{{\it Left)} SEM images of overhang discontinuity issue. {\it Middle)} Conformal sidewall deposition utilizing overhang etch, resulting in robust film continuity (see Section \ref{tes}). {\it Right)} SEM image of FIB-prepared cross-section of waterfall feature, showing conformal tungsten deposition. FIB image courtesy of Evans Analytical Group.}
    \label{wfall}
\end{figure}

Because it is deposited over an already etched structure, the TES film must maintain continuity while stepping down $\sim$330nm (the initial aluminum + tungsten layer) from the initial tungsten cap layer to the a-Si layer. Discontinuity in this region severs the phonon collection structure from the TES line (see Figures \ref{stzm}, \ref{wfallview}, and \ref{wfall}), preventing signal readout. To avoid this issue, the TES film must be a conformal layer closely following the topology, particularly the sidewall, of the trilayer structures. This region is designated as the ``waterfall'' region (see Figure \ref{wfall} for examples of this feature exhibiting both poor and good continuity). To prevent this problem, the ``overhang etch'' has been implemented into the process. This etch was tuned by performing many iterations of circuit fabrication on practice wafers with various overhang etch times using SEM imaging for feedback. Once established, the process was confirmed with thick substrates, again using SEM imaging.

\subsection{a-Si Etch and ``Trenching'' of iZIP Detectors}
\label{iZip}

\begin{figure}[ht]
    \centering
    \includegraphics[width=.57\textwidth]{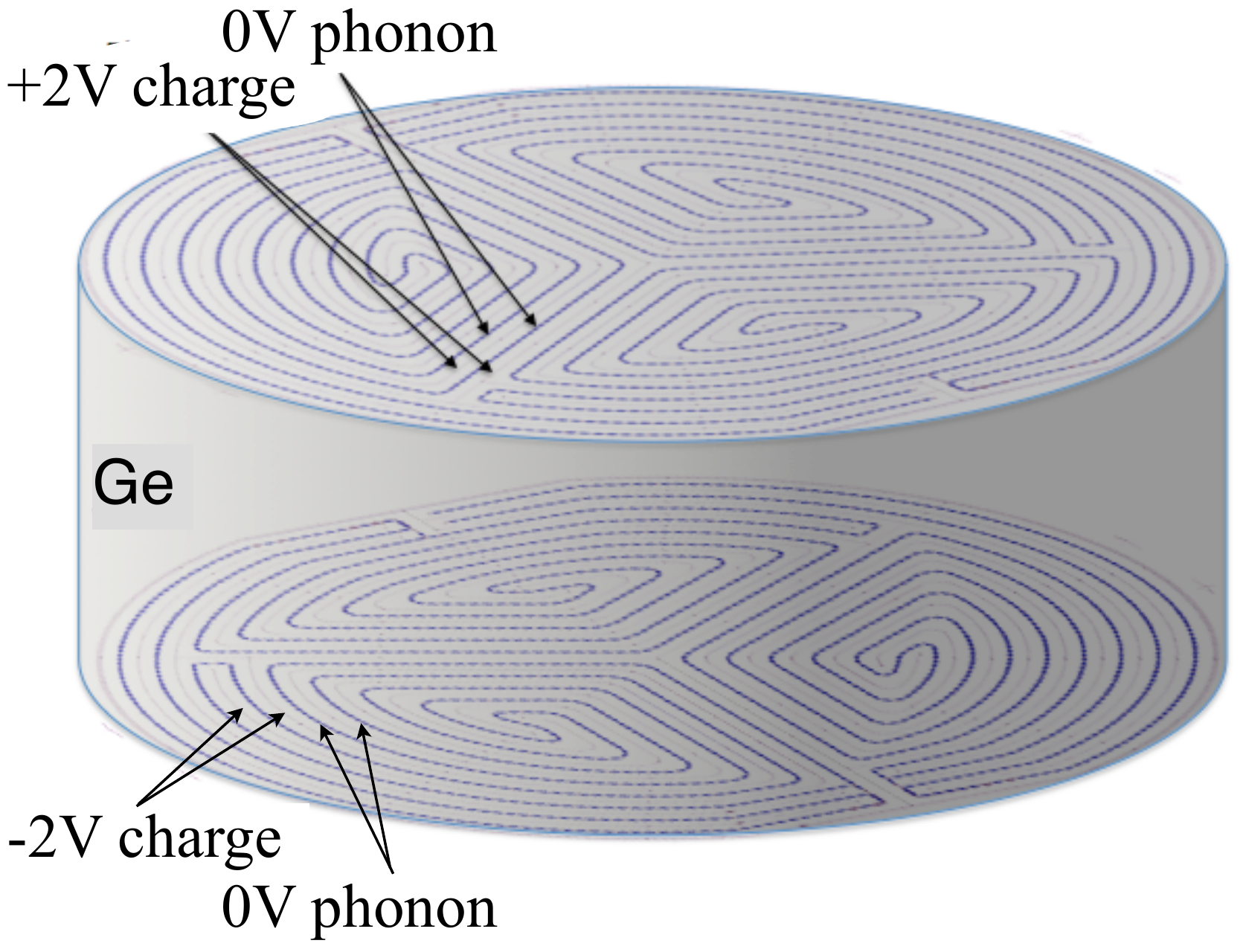}
    \includegraphics[width=.42\textwidth]{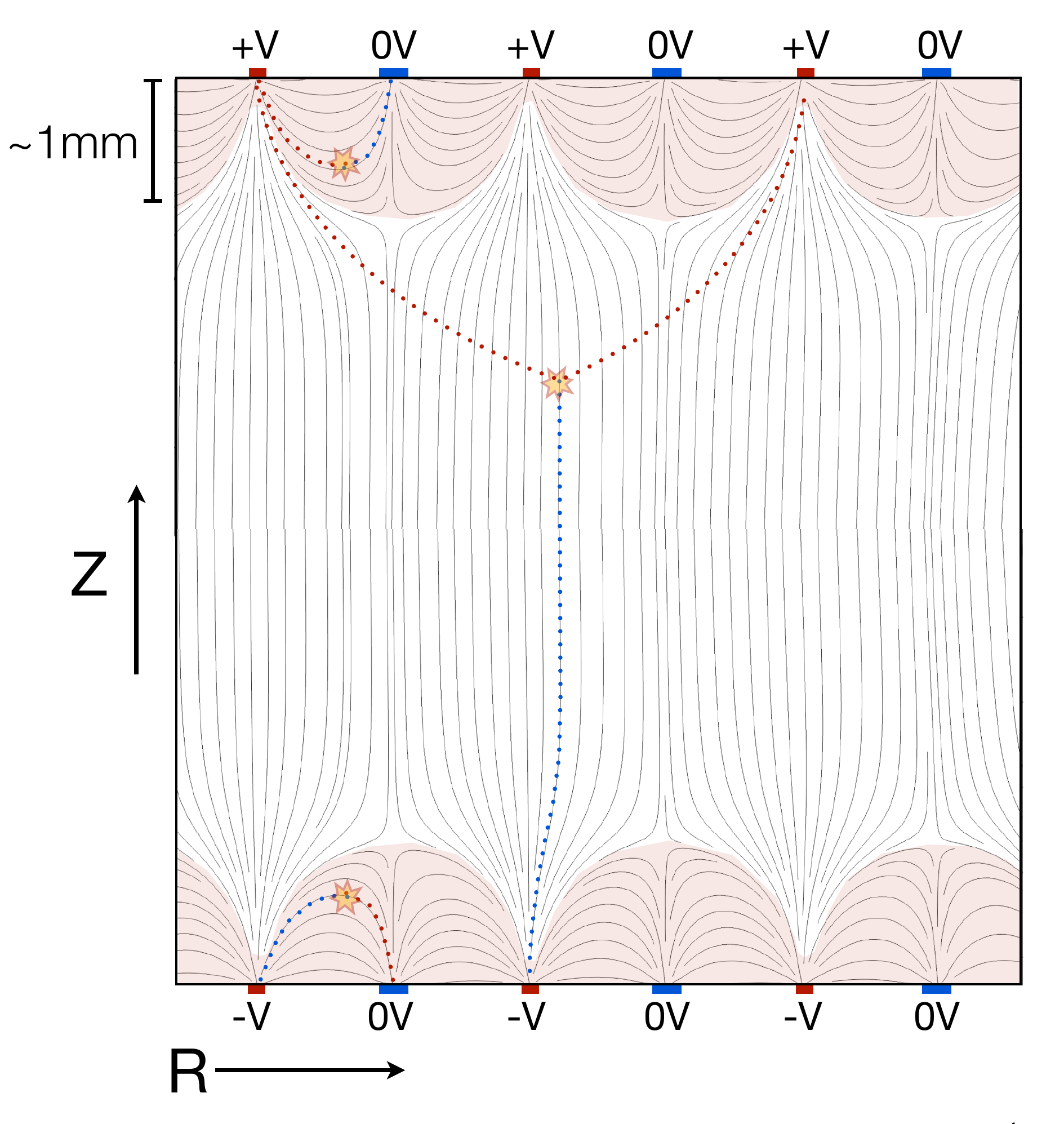}
    \caption{{\it Left)} Diagram of iZIP detector geometry/design used in SuperCDMS Soudan showing alternating biased charge collecting rails and 0V phonon rails (see Section \ref{iZip}). Figure from \cite{izip}. {\it Right)} Simulation of iZIP internal field lines, showing strong tangential electric fields at the surface and a uniform drift field in the detector bulk, a method proven to discriminate against the previously problematic surface events. Figure from \cite{hertel}.}
    \label{field}
\end{figure}

Discrimination between background and signal events using these detectors relies on a calibrated ratio of energy measured in the ionization channels versus that measured in the phonon channels. Charge carriers produced by events near the faces of a detector often fail to drift through the entire crystal to the appropriate electrode. This results in a reduced ionization collection signal, causing the event to be improperly identified in subsequent analyses.
A new circuit design has been implemented to combat this. The design is called the {\it i}ZIP ({\it interleaved} Z-sensitive Ionization and Phonon detector)\cite{pyle}\cite{izipa}\cite{izipb}, and as the name suggests, it utilizes interleaved electrodes on each surface. The interleaved electrodes alternate from ground to +2V on one face and from ground to -2V on the other (see Figure \ref{field}). This is in contrast to previous designs\cite{akerib} where one face is held at ground potential while the other is voltage biased. The interleaved design produces a very uniform field in the bulk but local regions of high field intensity near the surface. This causes the carriers (electrons and holes) produced near the surface to both be collected by the adjacent surface, with relatively little charge drifting to the opposite face. Therefore, any events with significant disparities in charge collection from one face to the other (i.e. failing the charge-symmetry requirement) are considered to be surface events. This procedure has been demonstrated to be very successful and is the design currently operating in SuperCDMS Soudan\cite{izip}. To fully realize the potential of this technique, detectors should be able to hold higher biases (producing stronger local surface fields) without breakdown. Limitations arise, however, as the electrode spacing is $\sim$1mm, and current begins to leak across the surface as voltage is increased, eventually resulting in breakdown. To reduce this problem and allow higher bias voltage, a trench is etched into the surface of the substrate, between the electrodes. For this purpose, the a-Si etch step is extended by $\sim$700\%. Because the gas used to etch the a-Si also etches the substrate material, this extra time allows etching of the substrate itself. The process has been tuned such that a trench of $\sim$1$\mu$m in depth is created between the electrodes and has been shown\cite{g10} to produce detectors that can hold much higher bias voltages without the problems mentioned previously (see Section \ref{results}).

\subsection{Photoresist Studies}
\label{pr}
\begin{figure}[ht]
    \centering
    \includegraphics[width=1\textwidth]{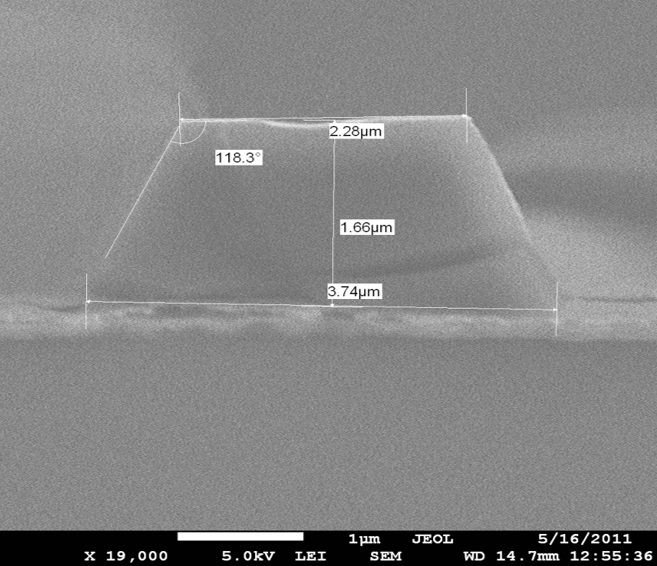}
    \caption{SEM image of a photoresist feature's cross-section. Due to improper UV exposure, this feature's sidewalls angle outwards, protecting a wider area of film from etching underneath, resulting in widening of circuit features. Note: This layer's thickness of 1.66$\mu$m corresponds to a previous spin coat recipe, using S1813 photoresist (see Section \ref{pr}).}
    \label{prsem}
\end{figure}

The tuning of the photoresist layer is of utmost importance as it dictates not only the geometries of the final detector circuit structure, but whether or not the deposited films survive the fabrication process at all. For this reason, much time was spent investigating the photoresist layer and photolithographic processing of this layer. The cross-section of the developed photoresist pattern is controlled with the UV exposure, which can result in angled sidewalls (inward or outward), changing the width of the film etched below (see Figure \ref{prsem}). A dedicated study of the UV exposure (varying UV power and time, with SEM feedback) was performed to prevent these problems from affecting our circuit features.

In previous CDMS detector designs, Shipley Microposit S1813 photoresist\cite{1800series} has been used for the photolithographic processing. This chemical was chosen for multiple reasons:
\begin{enumerate}
\item Resistance to etchant chemicals used in this process 
\item Ability to reproduce sub-micron line-widths
\item Viscosity to match our desired layer thickness ($\sim$1.4$\mu$m) with rotational speeds that produce optimum uniformity (3500-5500rpm \cite{1800series})
\item Compatibility with metal ion-free developers \cite{1800series}
\end{enumerate}
However, as substrates of larger mass joined the production line, it became desirable to decrease spin coating speeds (causing less strain on the spin coater and less risk to the substrates). For this reason, Shipley Microposit S1811\cite{1800series} is now used for its lower viscosity while still possessing the other characteristics mentioned above. Using S1811, the spin coat process is performed at 2300 rpm (as opposed to 4000 rpm required for S1813) for 60 seconds, producing a 1.4$\mu$m layer on each face. These parameters (along with those of the UV exposure mentioned previously) were tuned using feedback from SEM images confirming faithful reproduction of mask feature line-widths and robust cross-sections after exposure and development. 

Another photoresist issue that was studied and successfully remedied is that of a so-called ``edge bead''. After the spin coat process, a thick bead of photoresist can be seen around the edge of the substrate surface. Thicker than the nominal layer, this bead does not receive enough UV exposure and developing to be properly patterned/removed prior to etching. While there are no vital circuit features in this region, it prevents any films underneath from being etched away, leaving a metal band (which could potentially cause shorts) around the edge of the detector after the photoresist is removed. It is for this reason that the previously mentioned shadow mask is implemented, preventing deposition in this region (see Section \ref{segi}). This procedure has proven to be a low cost yet highly effective method of combating the edge bead problem, with negligible impact on detector patterning at radial extremities.

\section{Results to Date}
\label{results}

\begin{figure}[ht]
    \centering
    \includegraphics[width=1\textwidth]{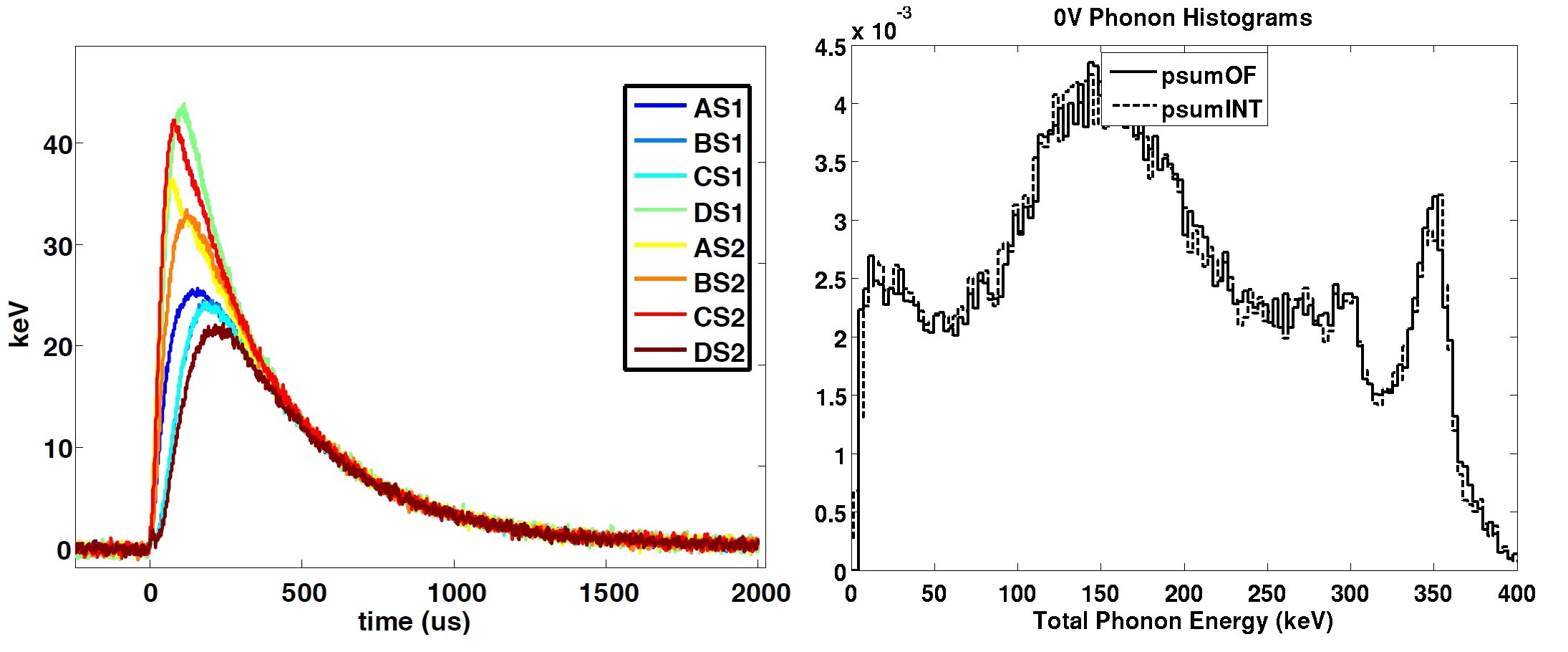}
    \caption{{\it Left)} Phonon pulses from detector G9F, fabricated at TAMU. {\it Right)} Calibration spectrum from detector G9F, clearly showing the 356 keV Ba$^{133}$ peak (see Section \ref{results}).}
    \label{g9}
\end{figure}

Using this process at the dedicated TAMU fabrication facility, detectors have been produced of the size and design of those in SuperCDMS Soudan. Test data from detector G9F, one of the first of these produced at TAMU, can be seen in Figure \ref{g9}, demonstrating pulses from operational phonon sensor channels as well as the 356 keV photopeak from a Ba-133 calibration source. In addition, this detector showed unparalleled TES T$_c$ uniformity {\it without} ion implant compensation (see Figure \ref{tcmap} and Table \ref{g9data}). However, it showed an inability to hold adequate bias voltage, leading to further tuning of the trenching process (see Section \ref{iZip}). The following detector, G10F (using the improved trenching process), demonstrated more than adequate ability to hold bias, showing no signs of leakage up to $\pm$5V (the limit of the test stand). Specification standards used to rate SuperCDMS detectors categorize this detector as ``very good''.  Subsequent testing showed functional charge performance up to 9V \cite{g10}, much higher than required for the experiment.  

\begin{figure}[ht]
    \centering
    \includegraphics[width=1\textwidth]{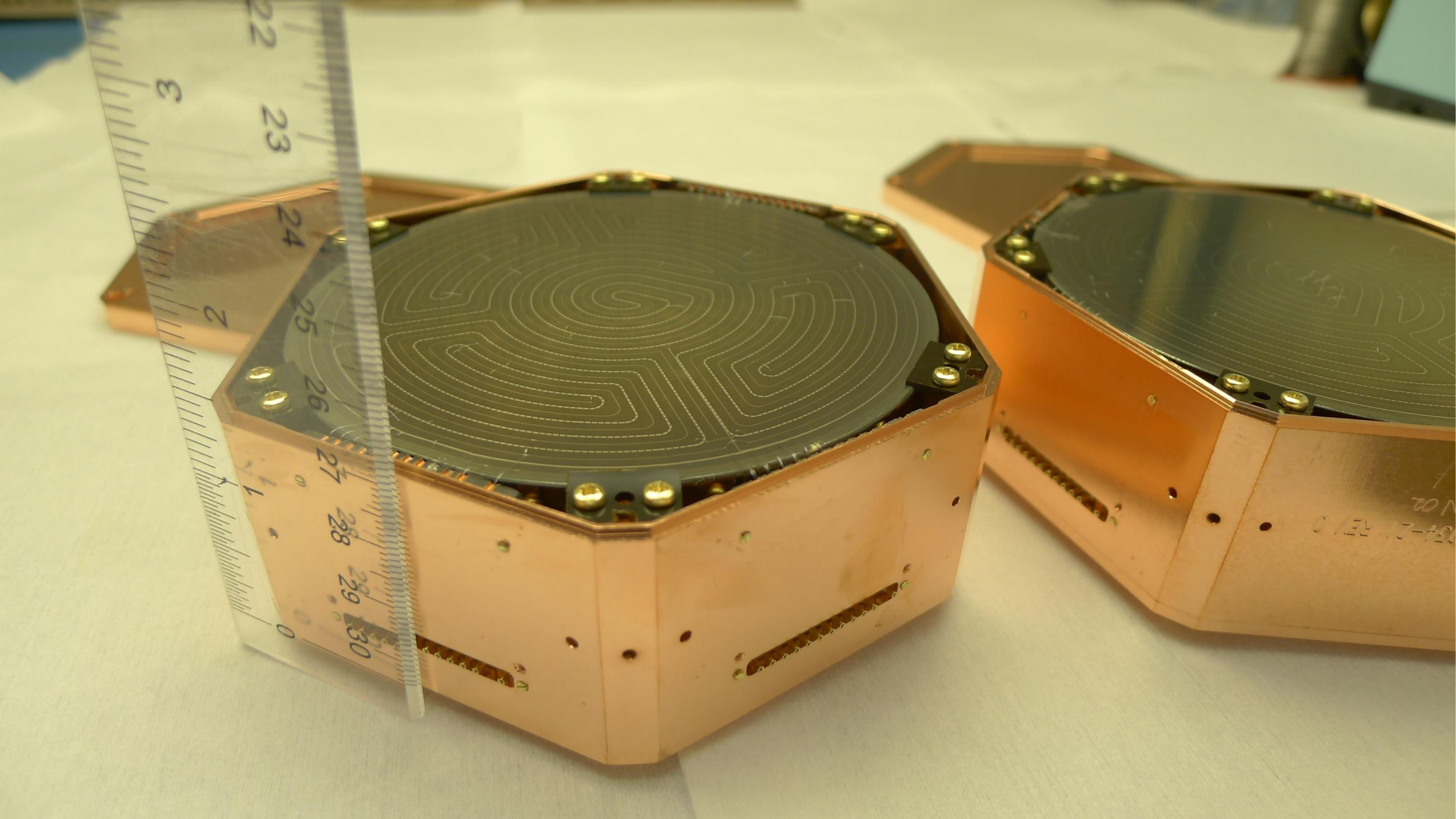}
    \caption{100mm x 33.3mm detectors fabricated at TAMU, currently awaiting testing.}
    \label{detector}
\end{figure}

Detectors produced at this facility have demonstrated performance that meets or exceeds the requirements for this experiment, certifying this location as an integral fabrication facility for SuperCDMS SNOLAB detectors. 100 mm x 33.3 mm thick science quality detectors were successfully produced at this facility in early 2013 (see Figure \ref{detector}) and are currently awaiting testing.

\section{Conclusion}

A robust, repeatable fabrication procedure has been established, demonstrated, and improved at the TAMU fabrication facility. Increased throughput as well as improvements made in the process itself are expected to contribute substantially to the success of the next generation SuperCDMS SNOLAB experiment. In particular, increased fabrication efficiency, improved TES T$_c$ consistency and uniformity, increased bias voltage ability due to substrate trenching, and improved signal collection from overhang studies will improve detector success rates, reducing fabrication {\it and} testing costs.

\section{Acknowledgments}
The authors would like to thank the CDMS team at SLAC/Stanford (especially Blas Cabrera, Paul Brink, Richard Partridge, Matt Cherry, Astrid Tomada and John Mark Kreikebaum), Sunil Golwala, Betty Young, Jae Woo Suh, the CDMS team at the University of Minnesota (especially Hassan Chagani, Priscilla Cushman, Matt Fritts, Tommy Hofer, Allison Kennedy, Vuk Mandic, Roxanne Radpour, David Strandberg, and Anthony Villano), and the University of California, Berkeley 75 $\mu$W dilution refrigerator test team. This work was supported by the Department of Energy (Contracts DE-FG02-13ER41918,  DE-AC03-76SF00098, DE-FG02-92ER40701, DE-FG02-94ER40823, and  DE-SC0004022), the National Science Foundation (Grant Nos. PHY-1102842 and NSF-0919599), and funding from Texas A\&M University.

\bibliographystyle{elsarticle-num}
\bibliography{mybibfile}

\end{document}